\documentclass[aps,amsfonts,prl,twocolumn,showpacs]{revtex4-1}

\usepackage{subfigure}
\usepackage{wrapfig}
\usepackage{textcomp}
\usepackage{graphicx}
\usepackage{amssymb}
\usepackage{amsmath}
\usepackage{bm}
\usepackage{amsmath, amsthm, amssymb}
\usepackage{hyperref}

\usepackage{amsfonts}
\usepackage{dcolumn}
\usepackage{float}
\usepackage{bm}

\def\beq{\begin{equation}}
\def\eeq{\end{equation}}
\def\bea{\begin{eqnarray}}
\def\eea{\end{eqnarray}}
\def\up{\uparrow}
\def\dn{\downarrow}

\begin{document}

\title{Non-equilibrium theory of tunneling into localized state in superconductor}

\author{Ivar Martin$^1$ and Dmitry Mozyrsky$^2$}

\affiliation{$^1$ Materials Science Division, Argonne National Laboratory, Argonne, Illinois 60439, USA\\
$^2$Theoretical Division, Los Alamos National Laboratory, Los Alamos, New Mexico 87545, USA
}

 \date{\today}

\begin{abstract}
A single static magnetic impurity in a fully-gapped superconductor leads to formation of an intragap quasiparticle bound state. At temperatures much below the superconducting transition, the energy relaxation and spin dephasing of the state are expected to be exponentially suppressed.  The presence of such a state can be detected in electron tunneling experiments as a pair of conductance peaks at positive and negative biases. Here we show, that for an arbitrarily weak tunneling strength, the peaks have to be symmetric with respect to the applied bias. This is in contrast to the standard result that the tunneling conductance is proportional to the local (in general particle-hole asymmetric) density of states. The asymmetry can be recovered is one allows for either a finite density of impurity states, or that impurities are coupled to another, non-superconducting, equilibrium bath.

\end{abstract}

\maketitle

{\em Introduction.} Conventional $s$-wave superconductors are remarkably robust with respect to nonmagnetic disorder \cite{Anderson}: potential scattering of electrons affects neither the superconducting gap, nor the transition temperature significantly. On the other hand, even weak magnetic impurities have been found to be strongly Cooper pair-breaking, leading to a rapid suppression of superconductivity \cite{AG}.

An exact treatment of a quantum magnetic impurity in a superconductor is a complex problem, which has only been solved numerically so far \cite{Sakai}. However, in the case when the magnetic moment can be treated as static, (approximately the case for atoms with large spin, $S$, or when conduction electrons only couple to one of the components of the spin), within the BCS approximation, the problem is easily solvable. The key result is the appearance of a localized, so called Yu-Shiba-Rusinov (YSR) quasiparticle state \cite{Yu, Shiba, Rusinov}. For finite density of impurities, these states fill the superconducting gap, eventually destroying superconductivity.

The presence of YSR-like states in superconductors has been confirmed by tunneling experiments \cite{tunnel, Yazdani} (see Fig. 1a).  The metal-insulator-superconductor junction experiment of Ref. \cite{tunnel} on Mn doped Pb revealed a $\sigma(V) =  dI/dV$ that is symmetric with respect to reversal of applied bias (particle-hole symmetry), with a clearly visible intra-gap peak whose energy remained approximately constant but the intensity grew with the increasing Mn concentration. Remarkably, the normal-tip STM experiment of Ref. \cite{Yazdani}, which allowed to look at individual magnetic ions of Mn or Gd on the surface of superconducting Nb, showed particle-hole asymmetric $\sigma(V)$. The asymmetry was attributed to the asymmetry in the particle and the hole content of the Bogoliubov quasiparticle associated with the YSR state. This however, raises a question why no such asymmetry had been observed in the earlier tunnel junction experiment \cite{tunnel}.

It is interesting to note that individual YSR states bear strong resemblance to the localized impurity, e.g. donor, states in semiconductors. Each donor or acceptor state in a semiconductor can be populated by at most two electrons (incuding spin). Consequently, if one were to perform a tunneling experiment
in a semiconductor, as long as the bias is insufficient to inject carries into conduction or valence band, the dc current will remain zero: after the tunneling electrons populate the initially unoccupied localized states, the current has to stop.
What makes the YSR states different? Just as in a semiconductor, the individual YSR states are infinitely sharp resonances, since there are no continuum states that they could hybridize with. Therefore, it would seem that continuous tunneling into YSR states should be impossible, in conflict wight the experimental observations. That YSR assumes classical impurity cannot be the issue, since even for a quantum impurity, the spectrum has only one bound quasiparticle state associated with every impurity \cite{Sakai}.
As we will show here,  the reason that the intra-gap tunneling through the localized states in a superconductor is possible lies in the ability of superconductor to violate particle conservation law: While it is impossible to introduce a single electron with subgap energy into bulk of superconductor, $two$ injected electrons with zero total energy can be absorbed by the condensate \cite{Andreev}.

\begin{figure}[htbp]
\begin{center}
\includegraphics[width=8.5cm]{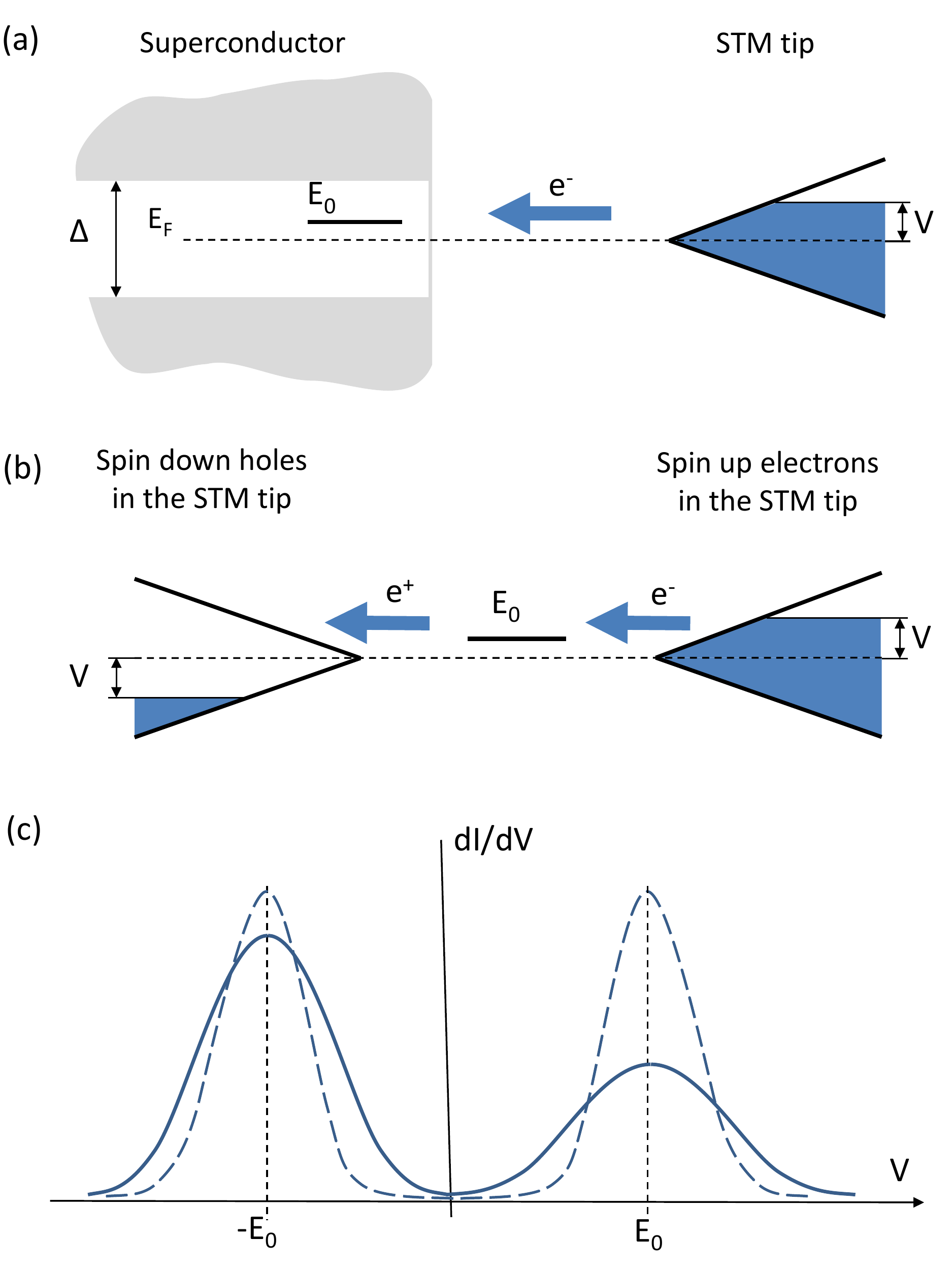}
\vspace{-0.1 cm}
\caption{\label{fig1} (a) Schematic representation of the problem: Electrons from an STM tip tunnel into a superconductor containing a single YSR state; (b) Effective representation after the particle-hole transformation on the spin-down tip electrons is performed; (c) Differential conductance of the system. The punctured line is the conductance of an ``ideal''system, i.e., when the broadening is caused by the couplig to the STM tip only. The solid line accounts for the ``extrinsic" broadening by an extra bath (other impurities or additinal normal reservoir).}
\end{center}
\end{figure}

This problem can be analyzed by means of non-equilibrium Green function formalism for superconductors \cite{Flensberg}. Here we will follow however a more physically transparent approach, valid in the case of singlet superconductors: By applying a partial particle-hole transformation, we convert the problem of tunneling from metallic tip to YSR state into the problem of tunneling between two non-superconducting spineless reservoirs through a single resonant level (Fig. 1b).  Each transfer of a spineless particle between the reservoirs in the equivalent model corresponds to the transfer of a pair of electrons between the metallic tip and the superconductor. The mapping allows to see immediately that for a single impurity $\sigma(V)$ has to be symmetric, regardless of the local particle-hole content of YSR state. The origin of this surprising result is that since in the absence of coupling to the tip YSR state has zero energy width, any arbitrarily weak perturbation can drive it out of equilibrium. The height of the peaks of $\sigma(V)$ is of the order of conductance quantum, $G_0=2 e^2/h$. In contrast, the standard approach to calculating the tunneling conductance assumes that the YSR remains in equilibrium with the superconductor, leading to the erroneous conclusion that for single magnetic impurity the tunneling conductance is simply proportional to the tunneling density of states \cite{Mahan}.

Why do some experiments show symmetric tunneling density of states \cite{tunnel}, and others don't \cite{Yazdani}?
The reason most likely lies in the broadening of the resonant level due to the presence of other nearby magnetic impurities, which allows electrons to tunnel into multiple YSR states simultaneously, or due to an additional relaxation channel for YSR states. The latter can be modeled as a metallic reservoir that remains in equilibrium with the superconductor and thus can easily absorb quasiparticles injected into the YSR state. We will explicitly consider here this possibility.

{\em Model.} The Hamiltonian for an $s$-wave superconductor with a magnetic impurity is \cite{Shiba}
\bea
H &=& H_{BCS} + H_{imp},\label{Shiba}\\
H_{BCS}& =& \int dr\Big [\sum_{\alpha}{\psi_{\alpha}(r) ^\dag \left(-\frac{\nabla^2}{2m} - \mu\right)\psi_{\alpha}(r)}\nonumber\\
&&  + \Delta_0 { \psi_{\up}(r)^\dag \psi_{\dn}(r)^\dag + \Delta_0 \psi_{\dn}(r) \psi_{\up}(r)}\Big],\\
H_{imp} &=& J S\big [ \psi_{\up}(0)^\dag \psi_{\up}(0)  -\psi_{\dn}(0)^\dag \psi_{\dn}(0)\big].
\eea
Here, $\psi_\alpha(r)$ is the annihilation operator for electron with spin $\alpha$ at location $r$, $m$ is the mass of electron, $\Delta_0$ is the unperturbed value of the superconducting order parameter (assumed real and positive for concreteness). For the impurity we assume a classical moment of size $S$ polarized in the positive $z$-direction (in the continuum limit value of the coupling constant $J$ is related to the atomic value by the factor of the unit cell volume, $a^3$).   This Hamiltonian can be diagonalized by the Bogoliubov quasiparticles \cite{deGennes}, $\gamma_n$, which satisfy $[H, \gamma_n^\dag] =E_n \gamma^\dag_n $ and can be expressed in terms of the electronic operators as

\beq
\gamma_n = \int dr[u_n(r)\psi_\up(r) + v_n(r)\psi_\dn^\dag (r)].
\eeq

The solution of the Bogoliubov equations for $u(r)$ and $v(r)$ reveals that static magnetic impurity leads to formation of localized state inside the superconducting gap \cite{Rusinov}, with the energy
\beq
E_0 = -\Delta_0 \,{\rm sign}(J)\,\frac{1-(\pi N_0J)^2}{1+(\pi N_0J)^2},
\eeq
and $(u,v)$ that oscillate with the Fermi wavevector and decay is space as $\exp(-r/\tilde\xi)/r$. The exponential decay is governed by the length $\tilde \xi = v_F/\sqrt{\Delta_0^2- E_0^2}$. Here $v_F$ is the Fermi velocity and $N_0$ is the normal state density of states in the superconductor. In general, $u(r)\ne v(r)$.

In addition to the localized states, there is a continuum of Bogoliubov's quasiparticles both for $E_n > \Delta_0$ and $E_n < -\Delta_0$. The Fermion operators can be expanded in terms of all Bogoliubov quasiparticles as $\psi_\up(r) = \sum_n u_n(r)\gamma_n$ and $\psi^\dag_\dn(r) = \sum_n v_n(r)\gamma_n$. Hence, the local density of electronic states is $N_\up(\omega) = \sum_n u_n^2(r)\delta(\omega - E_n)$ and $N_\dn(\omega) = \sum_n v_n^2(r)\delta(\omega + E_n)$. Note, that $single$ YSR level  contributes $two$ delta-functions at energies $\pm E_0$ with weights $u_0^2$ and $v_0^2$ that correspond to spin-up and spin-down states, respectively.

According to the standard theory of electron tunneling from a metallic contact \cite{Mahan}, at zero temperature the differential tunneling conductance $\sigma(V)$  is proportional to the density of states in the sample at $E = V$, which in the case of YSR states would correspond to, in general, asymmetric delta function peaks. However, as we discussed above, such treatment neglects the possibility of having non-equilibrium distribution function, which in fact, leads to a qualitatively different result.

The tunneling between atomically sharp tip and the sample can be described by the tunneling Hamiltonian,
\beq
H'= H_{tip} + t[d^\dagger_\sigma(r_0)\psi_\sigma(r_0) +\psi^\dag_\sigma(r_0)d_\sigma(r_0)],
\eeq
where $r_0$ corresponds to the location where the tip and sample wavefunctions overlap, with the matrix element $t$, and $H_{tip} = \sum_{k\sigma} (\epsilon_k^t - \mu^t) d_{k\sigma}^\dagger d_{k\sigma}$ is the Hamiltonian of the tip, with modes $d_k$.
The tunneling part of the Hamiltonian can be conveniently expressed in terms of the Bogoliubov quasiparticles. Since we are interested in the subgap conductance due to the YSR state, out of the full expansion we only need to keep terms related to it, $\psi_\up(r_0) \to  u_0(r_0)\gamma_0$ and $\psi^\dag_\dn(r_0) \to v_0(r_0)\gamma_0$. In the spin-down channel this leads to terms of the form $d_\dn^\dagger\gamma^\dag_0$, which do not conserve the number of particles.  A significant simplification occurs if one performs a particle-hole transformation of spin-down electrons in the tip, $\tilde d_\dn = d_\dn^\dagger$. For the spin down holes, $\epsilon_k^t \to -\epsilon_k^t$, $\mu^t \to -\mu^t$ (relative to the chemical potential of the superconductor), and the state occupation numbers $n_k \to 1-n_k$. In the new basis, the tunneling Hamiltonian becomes,
$$  t u(r_0)d^\dagger_\up(r_0)\gamma_0 - t v(r_0)\tilde d^\dagger_\dn(r_0)\gamma_0 + H.c.$$

The full transformed Hamiltonian, which includes the superconductor, the tip, and the tunneling between them, now conveniently conserves the number of particles. It corresponds to the problem  of tunneling of spinless particles between two reservoirs through a resonant level. The couplings to the two reservoirs are in general different due to the factors $u(r_0)$, $v(r_0)$. Schematically, the equivalent representation is illustrated in Figure 1b. The right reservoir correspond to spin-up electrons, and the left reservoir to spin-down holes. Notice that the process in which a particle is transferred from right reservoir to the left one, in terms of the original electrons corresponds to transferring two electrons (with spin up and spin down) into the superconductor, with the help of the YSR state. The initial and final energy of the spinless particle is the same; in the original language this corresponds to selecting two electrons with total energy equal to zero (relative to the superconductor's $\mu$).

The problem of tunneling through a resonant level is very well known \cite{BW}. The key quantities that enter are the tunneling rates between the level and the reservoirs, $\Gamma_1 = \pi N^t u_0^2(r_0)t^2$ and $\Gamma_2 = \pi N^t v_0^2(r_0)t^2$. The sum of these two rates determines the resonant level broadening. Interestingly, even when $\Gamma_1\ne\Gamma_2$, the particle current through the resonant level does not depend on the direction of bias, reaching the maximum value of ${(2e/\hbar)}\times 2\Gamma_1\Gamma_2/(\Gamma_1 + \Gamma_2)$ for large bias. The ratio of the current to the level width, measured in the voltage units, gives, up to a constant, the differential conductance. Since the magnitude of the current does not depend on the direction of bias, subgap $\sigma(V)$ is symmetric with respect to the sign of $V$.  With the numerical prefactors inluded, we find
\beq\label{sigmasym}
\sigma(\pm E_0)= \frac{2e^2}{h} \frac{4\Gamma_1\Gamma_2}{(\Gamma_1+ \Gamma_2)^2} = G_0 \frac {4u_0^2 v_0^2}{(u_0^2 + v_0^2)^2}.
\eeq
Thus the maximum value of conductance, which is achieved at the spatial locations $r$ where $u_0(r) = v_0(r)$ is equal to one quantum of conductance, and the spatial map of $\sigma(\pm E_0)$ can be used to determine the spatial dependence of the quasiparticle particle-hole content, $u_0(r)/v_0(r)$.

{\em Extra bath.} We now turn to the case when magnetic impurity is not fully isolated within superconductor. To allow for additional relaxation, we introduce a gapless metallic bath, whose chemical potential is pinned to the chemical potential of superconductor, into which YSR state can decay with rate $\Gamma_0$. If this rate is much faster than $\Gamma_{1,2}$, the YSR state will remain in equilibrium with superconductor, and we expect to recover the ``standard" result where $\sigma(V)$ is proportional to the density of states in superconductor.

We study this problem within the normal-state non-equilibrium Green function formalism. The current through the system is fully determined by the resonant level Green function \cite{MMH}, which in this case is
\bea
G^>(\omega)= -2i{\sum_{i=0, 1, 2}\Gamma_i[1-n_i(\omega)]\over (\omega - E_0)^2+(\Gamma_0+\Gamma_1 +\Gamma_2)^2 },\\
G^<(\omega)=2i{\sum_{i=0, 1, 2}\Gamma_i n_i(\omega)\over (\omega - E_0)^2+(\Gamma_0+\Gamma_1 +\Gamma_2)^2 },
\eea
with $n_{1(2)}(\omega)$ being the Fermi distribution functions for the reservoirs of spin up electrons and spin down holes, e.g. Fig. 1(b), $n_{1(2)}(\omega)=\{1+\exp{[(\omega\pm V)/T]}\}^{-1}$ and $n_0$ is the distribution function for the bulk of the superconductor, $n_0(\omega)=[1+\exp{(\omega/T)}]^{-1}$. The retarded (advanced) components are $G^{R(A)}=[\omega - E_0 \pm i(\Gamma_1+\Gamma_2+\Gamma_0)]^{-1}$. The current through the YSR level is given by
\bea
\nonumber
I(V) = {ie\over \hbar} \int {d\omega\over 2\pi} \{ (\Gamma_1-\Gamma_2)G^<(\omega) \ \ \ \ \ \ \ \ \ \ \ \ \ \ \ \ \ \ \\
+[\Gamma_1 n_1(\omega)-\Gamma_2 n_2(\omega)][G^R(\omega)-G^A(\omega)] \} ,
\eea
which is twice that of the case of a conventional resonant level \cite{BW}. 
The corresponding differential conductance $\sigma(V)=dI/dV$ at zero temperature has a simple two-Lorentzian form,
\bea
\sigma= 2G_0\left[{2\Gamma_1\Gamma_2+\Gamma_0\Gamma_1\over (V-E_0)^2+\Gamma_T^2}+{2\Gamma_1\Gamma_2+\Gamma_0\Gamma_2\over (V+E_0)^2+\Gamma_T^2} \right],
\eea
with $\Gamma_T=\Gamma_0+\Gamma_1+\Gamma_2$. If $\Gamma_0 \gg \Gamma_{1,2}$, the heights of the Lorentzian peaks at $\pm E_0$ are proportional to $u^2$ and $v^2$, respectively, wich is the standard density of states result (see Fig 1c, solid line).  Only when $\Gamma_0 \ll \Gamma_{1,2}$ that the symmetric $\sigma(V)$ is recovered, e.g., Eq.~(\ref{sigmasym}). Finite temperature does not change this conclusion. 

In view of this result, we conclude that in the STM experiment of Ref. \cite{Yazdani}, the impurity states cannot be considered to be isolated, i.e., their (unrelated to coupling to STM) linewidth was larger than the electron tunneling rate. On the other hand, the planar tunnel junction experiment of Ref. \cite{tunnel} showed symmetric $\sigma(V)$, indicating that the magnetic impurities were sufficiently diluted and decoupled form any extrinsic relaxation baths, so that the tunneling current could drive them out of equilibrium. We note here that since the the crossover from asymmetric to symmetric $\sigma(V)$ occurs when $\Gamma_0 \sim \Gamma_{1,2}$, varying $\Gamma_{1,2}$ in STM experiments by means of changing the tunneling distance and lateral tip location, can be used to determine the broadening $\Gamma_0$.

{\em Measurement of impurity spin.}  Spin-polarized tunneling into the YSR state allows to measure the impurity spin orientation. Upon impurity spin reversal, the Bogoliubov quasiparticles transform as $E_n \to -E_n$, and $(u_n,v_n) \to (v_n, -u_n)$. Spin-polarized STM tip can be modeled by assuming different densities of states for up and down electrons, $N_\up^t \ne N_\dn^t$. If impurity spin is up, then $\Gamma_{1\up} = \pi t^2 u_0^2(r_0)N_\up^t $ and $\Gamma_{2\up} = \pi t^2 v_0^2(r_0)N_\dn^t $; for impurity spin down,  $\Gamma_{1\dn} = \pi t^2 v_0^2(r_0)N_\up^t $ and $\Gamma_{2\dn} = \pi t^2 u_0^2(r_0)N_\dn^t $. Since $\Gamma_{i\up}\ne\Gamma_{i\dn}$ for $|u(r)|\ne |v(r)|$, the value of the current for the two impurity states will be different, and hence can be used to determine the spin orientation.

Thus, the presence of YSR state enables the measurement of the local moment orientation. However, as we will now show, it also leads to dephasing of the local moment. From the Hamiltonian (\ref{Shiba}), the effective magnetic field acting on the local moment is
\beq
{ h_z} = J [\psi^\dag_\up(0)\psi_\up(0) - \psi^\dag_\dn(0)\psi_\dn(0)].
\eeq
with the main contribution to the fluctuation of $h_z$ deriving from YSR state; the delocalized Bogoliubov quasiparticles can be neglected at low temperatures, as we will show below. That leaves
\beq
{ h_z} = J \left[u_0^2(0)^2 + v_0^2(0)\right]\gamma_0^\dag\gamma_0 - Jv_0^2(0).
\eeq
[Notably, within YSR approximation, the transverse field components are zero since they involve operator combinations $\gamma_0^2 = (\gamma_0^\dag)^2$]. The spin dephasing time $T_2$ is related to the fluctuations of this field as

$$\frac {1}{T_2} \sim S^2\int_{-\infty}^{\infty} dt \langle (h_z(t) -\langle {h_z}\rangle)(h_z(0) -\langle {h_z}\rangle )\rangle,$$
i.e.,  its determination reduces to evaluation of the zero-frequency correlation function of the YSR level occupation number.
The zero-frequency fluctuations of occupancy reach maximum in the sequential tunneling regime. These fluctuations can be easily determined from the classical rate equations to be $\Gamma_1\Gamma_2/(\Gamma_1 + \Gamma_2)^3$, which for the dephasing rate yields
$${\frac {1}{T_2}}_{seq} \sim \frac{J^2S^2 }{\Gamma} \left(\frac{a}{\xi_0}\right)^6.$$
(we assumed here that $\Gamma_1 \sim \Gamma_2 \equiv \Gamma$).
For instance, in the case of Nb the ratio of the coherence length to the lattice constant $\xi_0/a\sim 100$. Taking $J\sim 1e$V, and tunneling rate $\Gamma\sim 10^{10}$s$^{-1}$, which corresponds to the tunnel current of about 0.1 nA, the dephasing time is $10^{-8}$s.

In the low-bias regime, such that $|E_0|\gg (T, V) \gg \Gamma $, the fluctuations can be found using the same Green function formalism as we used to determine current. In this regime,
$${\frac {1}{T_{2}}}_{ l.b.} \sim \frac{\Gamma^3  \max(T, V)}{E_0^4} \, {\frac {1}{T_2}}_{seq},$$
which, for the same tunneling rate and $E_0$ of the order of $\Delta_0\sim 1me$V, gives $T_{2l.b.}\sim 10^{-4}$s. In this regime, the dephasing rate is proportional to $\Gamma^2$. That the contribution of the delocalized states in the superconductor to spin dephasing can be neglected, can be seen from the following qualitative argument. Let us consider each delocalized state in the same way as we did the YSR state. Since these states are delocalized, their broadening will scale as $u^2(0), v^2(0)\sim 1/V$. The number of these states is proportional to the sample volume  $V$, and hence their overall contribution will scale as $1/V$, vanishing for non-microscopic samples. Moving the tip away from the sample one can recover the dephasing and relaxation rates that are governed by thermal excitations, whose density is $\sim e^{-\Delta_0/T}$. This long dephasing rate makes localized spin states in superconductors an appealing framework for various quantum computing applications, including those based on Majorana fermions \cite{Yazd-Bern, ShibaRKKY}.


The results obtained here apply not only to YSR states, but to any other localized intragap states in superconductors, e.g. states in the vortex cores \cite{Caroli}, or to the case of normal-quantum dot-superconductor junctions \cite{QD, Flensberg}. In the case of quantum dots, the single particle states in the dot may provide the effective equilibrium reservoir that allows YSR level relaxation that we discussed above \cite{Fl2}. 

Experimentally it has been found that using a superconducting tip provides a way to sharpen the features associated with tunneling though the YSR state \cite{ShibaSC}. Theoretically, this problem can also be mapped onto tunneling of spinless particles between two reservoirs with energy dependent densities of states. Unlike in the normal tip case, however, the peaks that appear due to YSR states at $\pm(|\Delta_{tip}| + |E_0|)$ are no longer symmetric \cite{MM2} even in the absence of additional bath, consistent with experimental findings \cite{ShibaSC}.

{\em Acknowledgements.} We would like to thank E. Demler, J. Sau, A. Yazdani, A. Shnirman, and A. Koshelev for useful discussions. Work performed at Argonne National Laboratory (I.M.) is supported by the U. S. Department of Energy, Office of Science,
Office of Basic Energy Sciences, under Contract No. DE-AC02-06CH11357. Work at Los Alamos National Laboratory (D.M.) was carried out under the auspices of the NNSA of the U.S. Department of Energy  under Contract No. DE-AC52-
06NA25396.

%


\begin{thebibliography}{10}
\bibitem{Fl2} The theoretical interpretation of these experiments provided by Ref. \cite{Flensberg}, used technique different than ours, but also obtained symmetric $\sigma(V)$. They considered, however, only onsite tunneling into particle hole-symmetric YSR state (no potential scattering), where the density of states result would be symmetric as well.
\end{thebibliography}

\end{document}